\documentclass[a4paper,twoside]{article}

\usepackage{epsfig}
\usepackage{subcaption}
\usepackage{calc}
\usepackage{amssymb}
\usepackage{amstext}
\usepackage{amsmath}
\usepackage{amsthm}
\usepackage{multicol}
\usepackage{pslatex}
\usepackage{apalike}
\usepackage{algorithm2e}
\usepackage[bottom]{footmisc}
\usepackage{blindtext}
\graphicspath{{gfx/}}
\usepackage{hyperref}
\usepackage{tikz}
\usetikzlibrary{quantikz2}
\tikzset{slice/.append style={line width=1.5pt}}

\usepackage{subcaption}
\usepackage{verbatim}
\usepackage{url}
\usepackage{svg}
\usepackage{multirow}
\usepackage{subcaption}
\usepackage{graphicx}
\usepackage{tikz}
\usepackage{nicefrac}
\usepackage{orcidlink}

\usepackage{SCITEPRESS}     

\begin{document}

\title{Benchmarking Quantum Surrogate Models on Scarce and Noisy Data}


\author{\authorname{Jonas Stein\sup{1}\orcidlink{0000-0001-5727-9151}, Michael Poppel\sup{1}\orcidlink{0009-0005-1141-0974}, Philip Adamczyk\sup{1}, Ramona Fabry\sup{1}, Zixin Wu\sup{1}\orcidlink{0009-0006-4383-3127},\\Michael Kölle\sup{1}\orcidlink{0000-0002-8472-9944}, Jonas Nüßlein\sup{1}\orcidlink{0000-0001-7129-1237}, Daniëlle Schuman\sup{1}\orcidlink{0009-0000-0069-5517}, Philipp Altmann\sup{1}\orcidlink{0000-0003-1134-176X},\\Thomas Ehmer\sup{2}\orcidlink{0000-0002-4586-5361}, Vijay Narasimhan\sup{3}\orcidlink{0000-0002-7727-6860} and Claudia Linnhoff-Popien\sup{1}\orcidlink{0000-0001-6284-9286}}
\affiliation{\sup{1}LMU Munich, Germany}
\affiliation{\sup{2}Merck KGaA, Darmstadt, Germany}
\affiliation{\sup{3}EMD Electronics, San Jose, California}
\email{jonas.stein@ifi.lmu.de}
}


\abstract{Surrogate models are ubiquitously used in industry and academia to efficiently approximate given black box functions. As state-of-the-art methods from classical machine learning frequently struggle to solve this problem accurately for the often scarce and noisy data sets in practical applications, investigating novel approaches is of great interest. Motivated by recent theoretical results indicating that quantum neural networks (QNNs) have the potential to outperform their classical analogs in the presence of scarce and noisy data, we benchmark their qualitative performance for this scenario empirically. Our contribution displays the first application-centered approach of using QNNs as surrogate models on higher dimensional, real world data. When compared to a classical artificial neural network with a similar number of parameters, our QNN demonstrates significantly better results for noisy and scarce data, and thus motivates future work to explore this potential quantum advantage in surrogate modelling. Finally, we demonstrate the performance of current NISQ hardware experimentally and estimate the gate fidelities necessary to replicate our simulation results.
} 

\onecolumn \maketitle \normalsize \setcounter{footnote}{0} \vfill

\section{\uppercase{Introduction}}
\label{sec:introduction}
The development of new products in industrial innovation cycles can take dozens of years with R\&D costs ranging up to several billion dollars. At the center of such processes (which are ubiquitous to, e.g., chemical, construction, financial, materials and pharmaceutical industries) lies the problem of finding the outcome of an experiment, given a specific input \cite{doi:10.1080/713665669,WESTERMANN2019170,Batra2021,GALATA2021120338,https://doi.org/10.1002/cite.201800091}. Having to account for highly complex interactions in the examined elements typically demands for tedious real-world experiments, as computational simulations either have very long runtimes or only approximate the actual outcomes crudely. As a result, only a small number of configurations can be tested in each product development iteration, often leading to suboptimal and misguided steps. Additionally, experiments suffer from aleatoric uncertainty, i.e., imprecisions in the experiment set-up, read-out errors, or other types of noise.

A popular \emph{in silico} approach for accelerating such simulations are so called surrogate models, which aim to closely approximate the simulation model while being much cheaper to evaluate. The central element within the surrogate modeling approach is to fit a computational model to the available data. More recently, the usage of highly parameterized models like classical Artificial Neural Networks (ANNs) as surrogate models has gained increasing research interest, as they have shown promising performance in coping with the typically high dimensional solution space \cite{doi:10.1177/0954410019864485,10.3389/fenrg.2022.979168,Vazquez-Canteli_2019}. However, a substantial issue with using ANNs as surrogate models is overfitting, which is mainly caused by a combination of noisiness and small sample sizes of the available data points in practice. \cite{Stokes2020,doi:10.1021/ci500747n,https://doi.org/10.1002/minf.201501008,Ponzoni2017}

In contrast to classical ANNs, quantum neural networks (QNNs), have been shown to be quite robust with respect to noise and data scarcity \cite{https://doi.org/10.48550/arxiv.2209.05523,Mitarai_2018}. Furthermore, QNNs also natively allow to efficiently process data in higher dimensions than classically possible, displaying another possible quantum advantage.

Eager to investigate the usage of QNNs as surrogate models in a realistic setting (i.e., given a limited number of noisy data points), our core contributions amount to:
\begin{itemize}
    \item An exploration of the practical application of QNNs as surrogate functions beyond existing proofs of concept, i.e., by fitting scarce, noisy, data sets in higher dimensions.
    \item An extensive evaluation of the designed QNNs against a similarly sized ANN, constructed to solve the given tasks accurately when given many noise free data samples.
    \item An empirical analysis of NISQ hardware performance and an estimation of needed hardware error improvements to replicate the simulator results.
\end{itemize}

The remainder of this paper is structured as follows. Section~\ref{sec:background} lays out the theoretical background of (quantum) surrogate models. Section~\ref{sec:methodology} subsequently describes our approach, as well as the choice of benchmark data sets and the classical baseline. Section~\ref{sec:evaluation}  contains the results and discussion. Section~\ref{sec:conclusion} ultimately concludes our findings. 


\section{\uppercase{Background}}
\label{sec:background}
A surrogate function $g$ is typically used to approximate a given black box function $f$ for which some data points $\left(\left(x_1, f\left(x_1\right)\right), ..., \left(x_n, f\left(x_n\right)\right)\right)$ are given, or can be obtained from a costly evaluation of $f$. The goal in surrogate modelling is finding a suitable, efficient whitebox function $g$ such that
\begin{equation}
    d\left(f(x_i), g(x_i)\right) \leq \epsilon
\label{eq:surrogateeq}
\end{equation}
where $d$ denotes a suitable distance metric and $\epsilon>0$ is sufficiently small for all $x_i$. Possible employed functions and techniques for representing $g$ include polynomials, Gaussian Processes, radial basis functions and classical ANNs \cite{https://doi.org/10.1002/cite.201800091,QUEIPO20051,https://doi.org/10.1002/wics.73}.

In our approach, we use a QNN, which modifies the parameters $\theta$ of a parameterized quantum circuit (PQC) to approximate the function $f$ as described in \cite{Mitarai_2018,PhysRevA.103.032430}. Each input data sample $x_i$ is initially encoded into a quantum state $\ket{\psi_{\text{in}}(x_i)}$ and then manipulated by a series of unitary quantum gates of a PQC $U(\theta,x)$.
\begin{equation}
    \ket{\psi_{\text{out}}(x_i, \theta)} =  U(\theta,x)\ket{\psi_{\text{in}}(x_i)}
\end{equation}

Choosing a suitable measurement operator $M$ (e.g., the Pauli Z-operator for each qubit $\sigma^z_i$), the expectation value gets measured to obtain the predicted output data.
\begin{equation}
    g_{QNN}(x_i) = \bra{\psi_{\text{out}}(x_i, \theta)}M\ket{\psi_{\text{out}}(x_i, \theta)}
\end{equation}

Aggregating the deviation of the generated output from the original output data then provides a quantification of the quality of the prediction which will finally be used to update the parameters of the gates within the PQC in the next iteration.

To encode the provided input data sample, many possibilities such as basis encoding, angle encoding and amplitude encoding have been explored in literature \cite{9425837}. While some of these encodings, typically called feature maps, exploit the richer tool set available in quantum computing to very efficiently upload more than one classical data point into one qubit (e.g., amplitude encoding), others are less costly with the respect to the amount of state preparation gates required (e.g., angle encoding).

For the PQC, typically called \emph{ansatz}, many architectures have been proposed. These generally consist of parameterized single qubit rotation and entanglement layers. \cite{https://doi.org/10.1002/qute.201900070} provides an overview of various circuit architectures, together with their \emph{expressibility} and entangling capability. In this context, \emph{expressibility} describes the size of the subset of states that can be reached from a given input state by changing the parameters of the ansatz. The more states can be reached, the more universal the quantum function can be.

As shown in \cite{Schuld_2021_L}, quantum models can be written as partial Fourier series in the data, which can in turn represent universal function approximators for a rich enough frequency spectrum. Following this, techniques like parallel encoding, as well as data reuploading display potent tools in modelling more expressive QNNs \cite{Schuld_2021_L,P_rez_Salinas_2020}. More specifically, parallel encoding describes the usage of a quantum feature map in parallel, i.e., for multiple qubit registers at the same time, while data reuploading is defined as the repeated application of the feature map throughout the circuit.

The approximation quality achieved by the QNN (i.e., $\epsilon$ from equation \ref{eq:surrogateeq}) can be quantified by choosing a suitable distance function, such as the mean squared error. Employing a suitable parameter optimization method such as the \emph{parameter shift rule} allows for training the QNN. Notably, all known gradient based techniques for parameter optimization (such as the parameter shift rule) scale linearly in their runtime with respect to the number of parameters, limiting the number of parameters, that can be trained given a limited amount of time.~\cite{Mitarai_2018}


\section{\uppercase{Methodology}}
\label{sec:methodology}
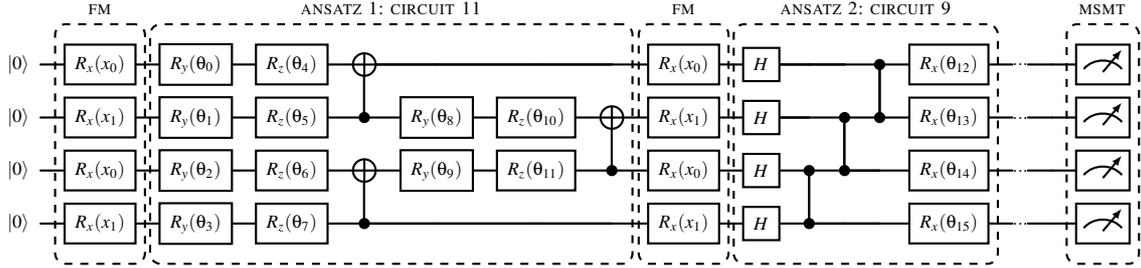
\begin {figure*}[htbp]
\centering
\begin{quantikz}[font=\scriptsize, column sep=9pt, row sep={20pt,between origins}]
\lstick{\ket{0}} & \gate{R_x(x_0)}\gategroup[wires=4,steps=1,style={dashed, rounded corners, inner xsep=0pt}, background]{{\sc fm}} & \gate{R_y(\theta_0)}\gategroup[wires=4,steps=6,style={dashed, rounded corners, inner xsep=0pt}, background]{{\sc ansatz 1: circuit 11}} & \gate{R_z(\theta_4)} & \targ{} & \qw & \qw & \qw & \gate{R_x(x_0)}\gategroup[wires=4,steps=1,style={dashed, rounded corners, inner xsep=0pt}, background]{{\sc fm}} & \gate{H}\gategroup[wires=4,steps=5,style={dashed, rounded corners, inner xsep=0pt}, background]{{\sc ansatz 2: circuit 9}} & \qw & \qw & \ctrl{1} & \gate{R_x(\theta_{12})} & ... & & \meter{}\gategroup[wires=4,steps=1,style={dashed, rounded corners, inner xsep=0pt}, background]{{\sc msmt}}  \\
\lstick{\ket{0}} & \gate{R_x(x_1)} & \gate{R_y(\theta_1)} & \gate{R_z(\theta_5)} & \ctrl{-1} & \gate{R_y(\theta_8)} & \gate{R_z(\theta_{10})} & \targ{} &  \gate{R_x(x_1)} & \gate{H} & \qw & \ctrl{1} & \ctrl{-1} & \gate{R_x(\theta_{13})} &... & & \meter{}  \\
\lstick{\ket{0}} & \gate{R_x(x_0)} & \gate{R_y(\theta_2)} & \gate{R_z(\theta_6)} & \targ{1} & \gate{R_y(\theta_9)} & \gate{R_z(\theta_{11})} & \ctrl{-1} &  \gate{R_x(x_0)} & \gate{H} & \ctrl{1} & \ctrl{-1} &\qw & \gate{R_x(\theta_{14})} &... & & \meter{}  \\
\lstick{\ket{0}} & \gate{R_x(x_1)} & \gate{R_y(\theta_3)} & \gate{R_z(\theta_7)} & \ctrl{-1} & \qw & \qw & \qw &  \gate{R_x(x_1)} & \gate{H} & \ctrl{-1} & \qw & \qw & \gate{R_x(\theta_{15})} &... & & \meter{}  
\end{quantikz}
\caption{The general QNN architecture used in this paper, exemplarily showing two layers, each comprised of a data encoding and a parameterized layer. It combines data reuploading by inserting a feature map (``FM") in each layer with parallel encoding by using two qubits per input data point dimension. For the parameterized part of the circuit, two different circuits from established literature have been combined: Circuit 11 is alternating with circuit 9 to create the required minimum number of parameters \cite{https://doi.org/10.1002/qute.201900070}. CNOT, Hadamard and CZ gates create superposition and entanglement, while trainable parameters $\theta_i$ allow the approximation of the surrogate model. After repeated layers, the standard measurement (denoted with ``MSMT") is applied to all qubits.}
\label{fig_vqc_example}
\end{figure*}    

To benchmark the practical application of QNNs as surrogate functions, we propose the following straightforward procedure: (1) Identify suitable QNN architectures, (2) select a realistic data set and (3) choose a reasonable classical ANN as baseline.

\subsection{Identifying suitable ansätze}
Following the description of all QNN components in section \ref{sec:background}, we now (1) identify a suitable encoding, (2) select an efficient ansatz for the parameterized circuit, then (3) combine both using layering and finally (4) choose an appropriate decoding of measurement results to represent the prediction result. 




Paying tribute to the limited hardware capabilities of current NISQ hardware, we employ angle encoding and hence focus on small- to medium-dimensional data sets. In particular, angle encoding has the useful property of generating desired expressibility while keeping the required number of gates and parameters low, resulting in shallower, wider circuits compared to more space-efficient encodings such as amplitude encoding \cite{Ostaszewski_2021,Mantri2017}. A possible implementation can be seen in the first two wires of figure \ref{fig_vqc_example}, where two-dimensional input data $x=\left(x_0, x_1\right)$ is angle encoded using $R_x$ rotations. Notably, additional normalization of the entries $x_i$ to an interval inside $\left[0,2\pi \right[$ has been proven useful~\cite{Koelle23}.

For selecting an efficient ansatz, we use established circuits proposed in literature \cite{https://doi.org/10.1002/qute.201900070}, that allow for a high degree of expressibility and entanglement capacity and thus harnessing essential tools allowing for quantum advantage. An important feature of these ansätze is, that they underlie an intrinsic structure, which allows scaling their width to fit an arbitrary number of qubits. Furthermore, we acknowledge a core result of ansatz design, i.e., the number of parameters in the ansatz should at least be twice the number of qubit encodings in the entire circuit to fully exploit the spectrum of the necessary non-linear effects generated by data reuploading and parallel encoding \cite{Schuld_2021_L}. Targeting our circuit choice towards an optimal trade-off between high expressibility and a low number of parameters (while adhering to the above mentioned lower bound), we use a combination of the two most expressive circuits from \cite{https://doi.org/10.1002/qute.201900070}, i.e., ``circuit 9" and ``circuit 11", as their concatenation perfectly matches the lower bound of parameters. For details, see figure~\ref{fig_vqc_example}.

Apart from the employed layered circuit architecture, we also employ parallel encoding and data reuploading to allow building a sufficient quantum function approximator, in-line with \cite{Schuld_2021_L} and most PQCs employed in literature. While data reuploading and parallel encoding both have a similar effect on expressibility \cite{P_rez_Salinas_2020,PhysRevA.103.032430,schuld2021machine}, in preliminary experiments conducted for this paper data reuploading appeared to be more effective than parallel encoding, as it led to faster evaluations with higher R2 scores for our data sets. Additional empirical results led to the final circuit architecture displayed in figure \ref{fig_vqc_example}, where a small amount of parallel encoding was applied (i.e., only once), while the focus is on data reuploading. In practice, empirical data shows, that the number of necessary layers can vary from data set to data set by a lot (in our cases, from 6 layers up to 42 layers), for details, see section~\ref{sec:evaluation}.

After the various data encoding and parameterized layers in the QNN, the expectation value of the qubits must be calculated with some measurement operator in order to extract classical information out of the quantum circuit. Having tried various different possibilities (tensor products of $\sigma_z$ and $I$ matrices), preliminary studies to this work led to the selection of the standard measurement operator $\sigma_z$ for all qubits involved. Note that this demands for a suitable rescaling of the output values if the image of the function to be fitted lives outside the interval $\left[ -1, 1\right]$ (which typically is a parameter that can be estimated by domain experts in the given use case). The difference between the obtained output and the original data is calculated as the mean squared error, and the parameters $\theta$ for the next iteration are generated with the help of an optimizer. Following the optimizer benchmark study results provided in \cite{Joshi_2021,9345015}, \emph{COBYLA} \cite{COBYLA_1} was chosen for this purpose.

\subsection{Selecting benchmark data sets}
\label{subsec:datasets}
In previous work, \cite{Mitarai_2018} and \cite{Qiskit} showed that QNNs can fit very simple functions with one-dimensional input like $sin(x)$, $x^2$ or $e^x$. To go beyond these proofs of concept, we expand this to the following, more complex, standard, $d$-dimensional benchmark functions regularly used for benchmarking data fitting:
\begin{itemize}
    \item \emph{Griewank} func. $ \sum_{i=1}^d \nicefrac{x_{i}^2}{4000}-\prod_{i=1}^d\cos\left(\nicefrac{x_i}{\sqrt{i}}\right)+1$ for $x\in\left[-5, 5\right]^d$ \cite{GRIEWANK}
    \item \emph{Schwefel} func. $418.9829 d - \sum_{i=1}^d x_i \sin\left(\sqrt{\left|x_i\right|}\right)$ for $x\in\left[-50, 50\right]^d$ \cite{SCHWEFEL}
    \item \emph{Styblinski-Tang} func. $\nicefrac{1}{2}\sum_{i=1}^d \left(x_{i}^4 - 16x_{i}^2+5x_i\right)$ for $x\in\left[-5, 5\right]^d$ \cite{StyblinskiTang}
\end{itemize}
For each of these benchmark functions, we used two-dimensional input data, which, together with the one-dimensional function value, can still be presented in three-dimensional surface plots. In addition to that, we also investigate the performance on the real world data set \emph{Color bob} \cite{H_se_2021}, which contains 241 six-dimensional data points from a chemical process related to colorimetry. The intervals of the functions are chosen to balance complexity and visual observability, while keeping the number of required parameters in the PQC at a level that still allows for the necessary iterative calculations within reasonable time (i.e., a couple of days on our hardware). To facilitate an unambiguous angle encoding (for details, see \cite{Koelle23}), we normalize the data to the range $\left[0, 1\right]$ in all dimensions. Note that values between $0$ and $2\pi$ would have also been feasible for angle encoding, but the interval $\left[0, 1\right]$ showed even better performance in our experiments.

\subsection{Selecting a classical baseline}
\label{subsec:classicalANN}
As classical machine learning is tremendously better understood and currently far more performant than quantum machine learning, it is strongly to be assumed, that ANNs can be identified, that achieve (near) perfect results in all of our experiments, as the datasets are comparably small and simple. Aiming to provide a meaningful comparison nevertheless, we take a standard approach of choosing a similarly sized ANN as baseline, i.e., in terms of the number of parameters. Preliminary studies conducted for this article show, that already quite small ANNs can achieve accurate fits, i.e., a fully connected feedforward neural network consisting of an input layer (with as many neurons as dimensions in the domain space of the function), two hidden layers (containing 7 neurons and using a sigmoid activation function) and an output layer of size one. While the in- and output layers are fixed in size by the datasets used, the number of $n=7$ neurons per hidden layer is an empirical result obtained by iteratively increasing $n$, until the R2 score exceeded 90\% for every dataset. This amounts to 140 parameters, which corresponds to a QNN comprised of 9 layers of the employed architecture (as displayed in figure \ref{fig_vqc_example}). As no regularization technique was applied for the QNN, none is used for the ANN either, in order to compare the two models on a like-for-like basis. Analog to the QNN, we employ the mean squared error loss. For parameter training, we use the popular ADAM optimizer \cite{Adam_Opt} with 50,000 iterations.

\section{\uppercase{Evaluation}}
\label{sec:evaluation}

Having prepared a suitable QNN architecture, a classical ANN as baseline and a variety of challenging benchmark functions, we now evaluate the applicability of QNNs as surrogate functions in domains with scarce and noisy data. For all experiments, a 80/20 train/test split was used, and all displayed results show the performance on the test data.

\subsection{Baseline results for noiseless data}
To quantify our results, we use the \emph{R2 score}, a standard tool to measure the similarity of estimated function values to the original data point values, as well as visual inspection, to assess how well the shape of the original function gets approximated. The R2 score is defined such that it yields 1 if the model perfectly predicts the outcome, and lower values, the less well it predicts the outcome (where random guessing amounts to an R2 score of 0):
\begin{equation}
R2(y,\hat{y}) = 1 - \frac{\sum\limits^n_{i=1} (y_i - \hat{y_i})^2}{\sum\limits^n_{i=1} (y_i - \bar{y_i})^2}
\label{eq:R2score}
\end{equation}
Here, $n$ is the number of given samples, $y_i$ is the original value, $\hat{y_i}$ is the predicted value and $\bar{y_i} = \nicefrac{1}{n}\sum_{i=1}^n y_i$.

\begin{figure}
  \begin{subfigure}{0.49\columnwidth}
  \includegraphics[width=\textwidth]{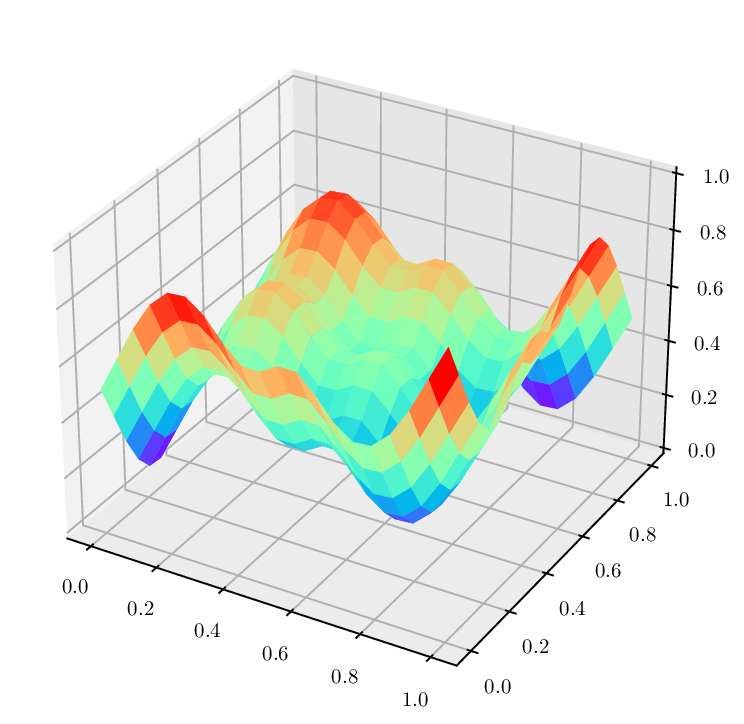}
  \caption{Original Schwefel function surface.}
  \label{subfig:clean_2dSchwefel}
  \end{subfigure}
  \hfill
  \begin{subfigure}{0.49\columnwidth}
  \includegraphics[width=\textwidth]{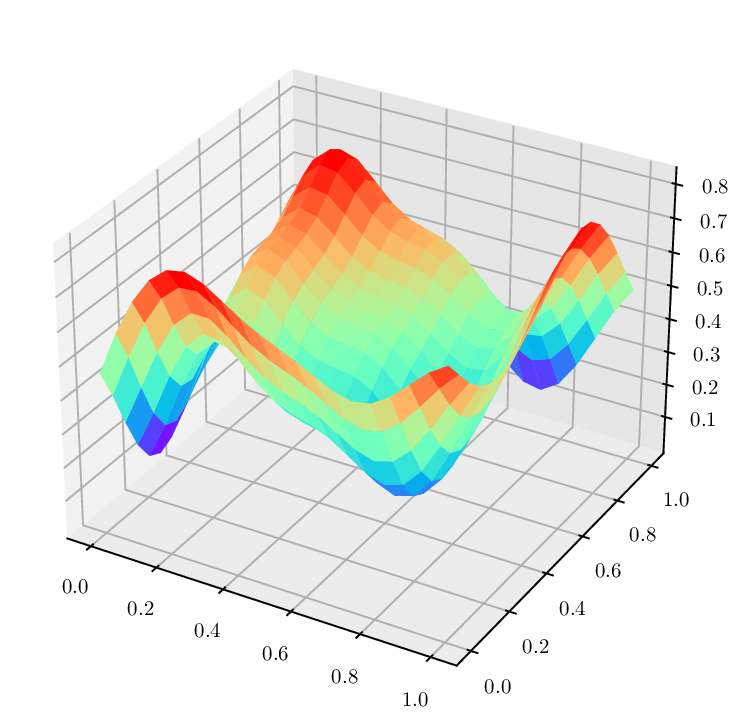}
  \caption{QNN generated surface, R2 score of 0.94.} 
  \label{subfig:QNN_Schwefel}
  \end{subfigure}
  \caption{Qualitative performance of the quantum surrogate for the Schwefel function with two-dimensional input.}
  \label{fig:Schwefel}
\end{figure}


For all benchmark functions with two-dimensional input data mentioned in section~\ref{subsec:datasets}, the QNN achieves very good R2 scores, even obtaining values above 0.9, when given the "full dataset" (i.e., an evenly spaced 50-by-50 grid of 2500 noiseless data points, as described in section~\ref{subsec:noiseandscarcity}). To achieve this solution quality, we identify a minimum of 20 (42) layers and 3000 (4000) optimization iterations required for the Griewank (Schwefel and Styblinski-Tang) functions. Exemplary results of the original surface and the surrogate model surface for the two-dimensional Schwefel function are displayed in figure \ref{fig:Schwefel}. For the \emph{Color bob} data set, we obtain an R2 score close to 0.9 with 3 layers consisting only of the feature map and ansatz 1 from figure~\ref{fig_vqc_example}, with merely 500 optimization iterations given the full dataset. To account for the increased dimensionality in the input data, we used five qubits, and skipped the parallel encoding, as the results turned out to be good enough without it. This demonstrates the first evidence towards promising performance of quantum surrogate models on industrially relevant, real world, dataset for very minor hardware requirements.

\subsection{Introducing noise and data scarcity}\label{subsec:noiseandscarcity}
In real world applications, the available data samples are often scarce and noisy. In order to model this situation, we introduce different degrees of noise on varying training set sizes. For this rather time intensive part of the evaluation, we focus on the Griewank function, as the number of layers and iterations (i.e., the computational effort) required to find a suitable quantum surrogate model for this function is a lot lower than for the others.

Given the discussed QNN (see figure \ref{fig_vqc_example}) and classical ANN (see section \ref{subsec:classicalANN}), we evaluate them for a range of 100, 400, 900, 1600 and 2500 data points (selected analog to selecting data points in grid search) while also introducing standard Gaussian noise factors \cite{Truax1980HandbookFA} of 0.1, 0.2, 0.3, 0.4 and 0.5 on the input data. More specifically, the noise is applied using the following map $\left(x_i,f(x_i)\right)\mapsto\left(x_i,f(x_i) + \delta \nu\right)$, where $\delta$ denotes the noise factor and $\nu$ is random value sampled from a Gaussian standard normal distribution $\varphi (z) = \nicefrac{\exp\left(\nicefrac{-z^2 }{2}\right)}{\sqrt{2\pi}}$. Adding noise on the input data can therefore be thought of as, an imprecise black-box function execution (e.g., due to inexact measurements in chemical experiments). For simplicity, we use the term \emph{sample size} in the following to denote the granularity of the grid in both dimensions, i.e., we investigate sample sizes of 10, 20, 30, 40, and 50.

\begin{figure}[!t]
\centering
\includegraphics[width=\columnwidth]{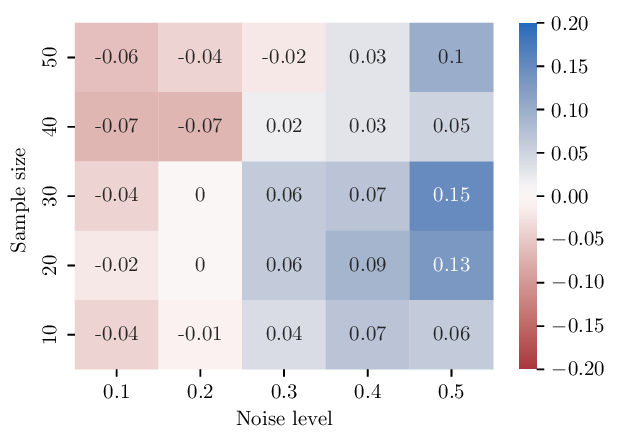}
\caption{Delta\_R2 score obtained by subtracting the classical ANN R2 score from the QNN R2 score for different noise levels ($x$-axis) and sample sizes ($y$-axis) for the Griewank function with two-dimensional input data. Positive values indicate a performance advantage of the QNN (as can be seen for higher noise levels and smaller simple sizes), while negative values represent a disadvantage of the QNN.}
\label{fig_heatmap_coldwarm}
\end{figure}

As we are mainly interested in the relative performance of the QNN and ANN against each other, we focus our evaluation on the difference of their R2 scores $\text{Delta\_R2}=R2_{\text{QNN}}(y,\hat{y})-R2_{\text{ANN}}(y,\hat{y})$, as depicted in figure \ref{fig_heatmap_coldwarm}. The results show that our QNN has a tendency to achieve better R2 scores for higher noise levels compared to the classical ANN. This trend seems to be even more obvious for smaller sample sizes compared to larger ones. This indicates, that QNNs can have better generalization learning ability when given scarce and noisy data. 

\begin{figure}
  \begin{subfigure}{0.49\columnwidth}
  \includegraphics[width=\textwidth]{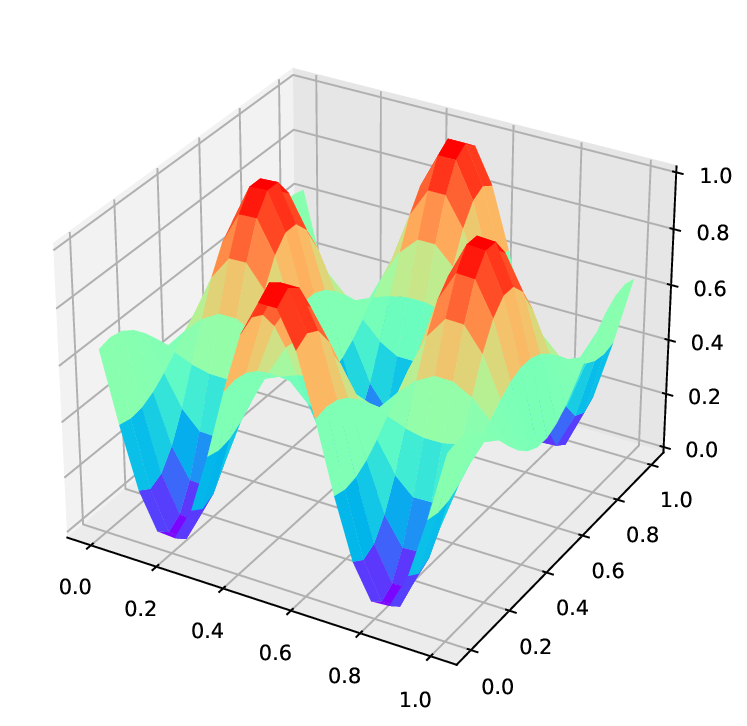}
  \caption{3-D Griewank surface.}
  \label{subfig:clean_2dGriewank}
  \end{subfigure}
  \hfill
  \begin{subfigure}{0.49\columnwidth}
  \includegraphics[width=\textwidth]{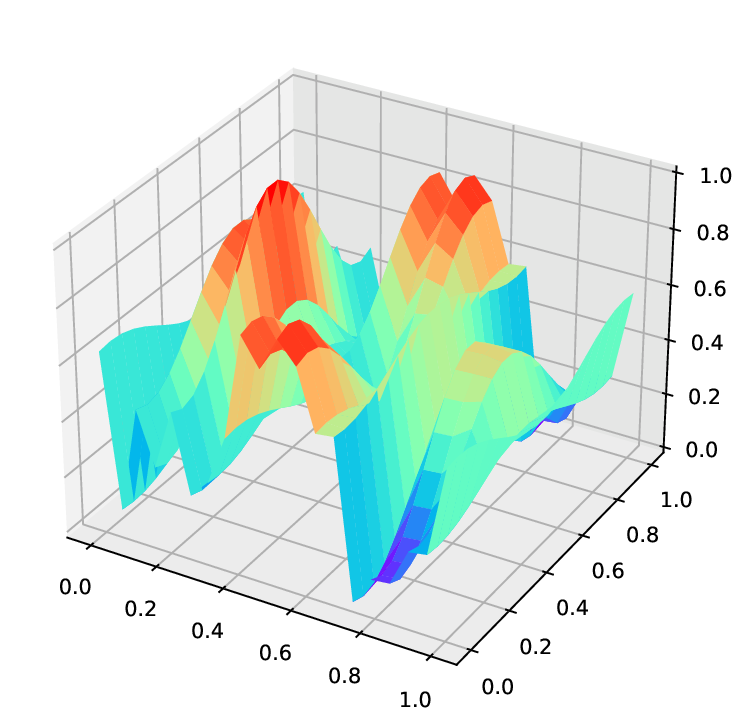}
  \caption{Noisy surface.\vspace{0.05cm}} 
  \label{subfig:noisy_2dGriewank}
  \end{subfigure} 
  \begin{subfigure}{0.49\columnwidth} 
  \includegraphics[width=\textwidth]{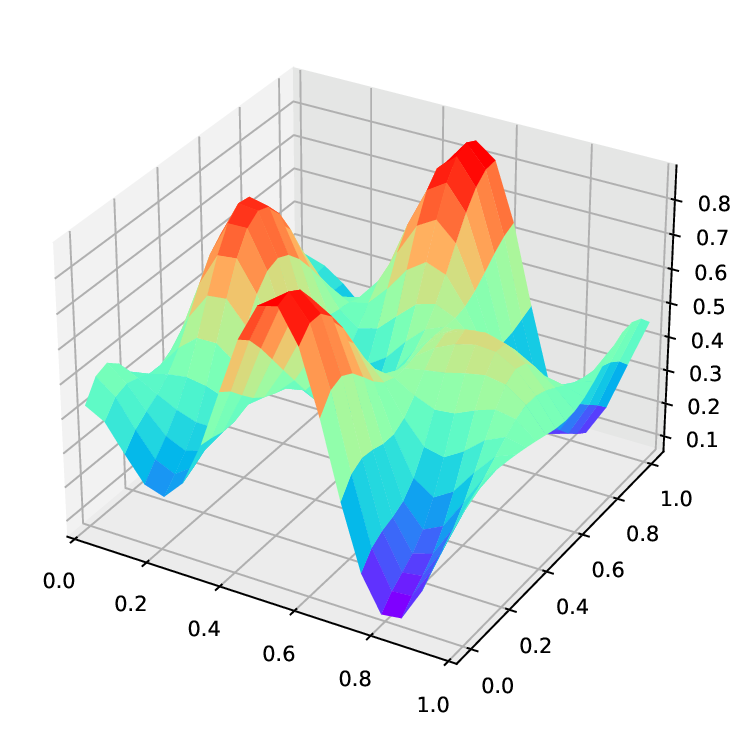} 
  \caption{Surface-fit of the QNN.} 
  \label{subfig:QNN_fitted}
  \end{subfigure}  
  \hfill 
  \begin{subfigure}{0.49\columnwidth} 
  \includegraphics[width=\textwidth]{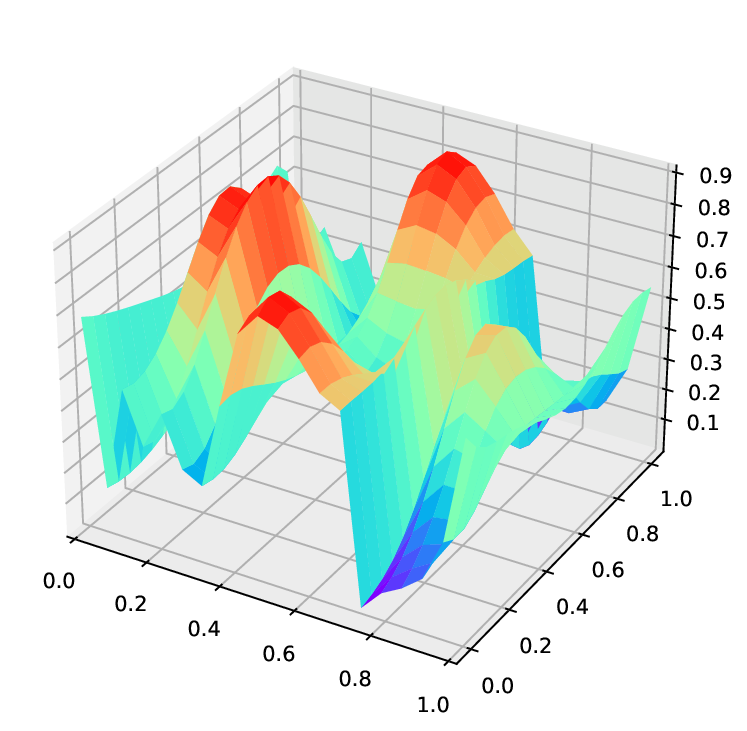} 
  \caption{Surface-fit of the ANN.} 
  \label{subfig:CNN_fitted}
  \end{subfigure}
  \caption{Surface plots of the Griewank function when introducing noise and sample scarcity. The input to the QNN and classical ANN was modified by multiplication of standard-normal noise with a factor of 0.5 on the 400 individual input data points (which corresponds to a sample size of 20).}
  \label{fig_6surfaces}
\end{figure}

In order to also examine the results generated by the QNN and the classical ANN visually, we depict plots of the surrogate model for the Griewank function as well as the original Griewank function and the function resulting from overlaying noise with a factor of 0.5 in figure \ref{fig_6surfaces}. The QNN shows more resilience than the ANN for increasing noise levels and clearly maintains the shape of the original function a lot better. Notably, for noise levels below 0.3, the classical ANN was able to achieve higher R2 scores than the QNN. However, overall, the better performance of QNN for noisy and scarce data points towards its comparatively better generalization capabilities.

\subsection{NISQ Hardware Results}
Today’s HPC quantum circuit simulators have shown the capability to simulate small circuits up to 48 qubits \cite{10.1145/3295500.3356155}. Taking this as an approximate bound for how many qubits a circuit can consist of in analytic calculations, it becomes apparent, that classical simulations face a clear limit when assessing high dimensional input data when using qubit intensive data encoding. Based on our experiments, two to three qubits per dimension showed the best results. When facing a small scale data set with industrial relevance, e.g., the 29-dimensional input data as available in PharmKG, a project of the pharmaceutical industry \cite{10.1093/bib/bbac404}, this already would require 58 to 87 qubits, thus quickly exceeding what is technically possible with classical computers. Therefore, one needs to rely on quantum hardware in order to examine scaling performance for higher dimensional data sets, optimally using fault tolerant qubits.

\begin{figure}[!t]
\centering
\includegraphics[width=\columnwidth]{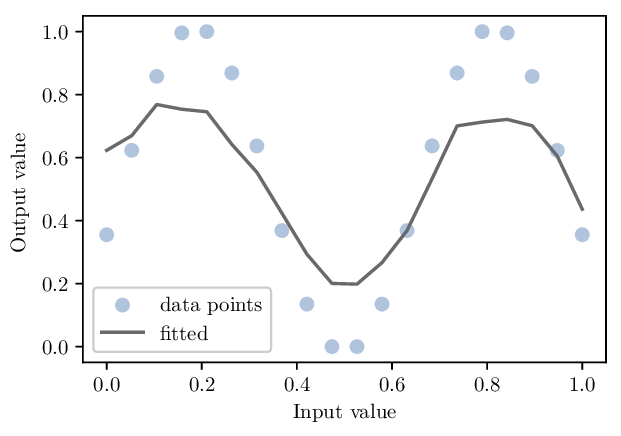}
\caption{Original data points (sample size of 20) for a Griewank function with one-dimensional input data and the quantum surrogate function that has been obtained by running 100 iterations of 6 layers of our ansatz described in figure~\ref{fig_vqc_example} on the ibmq\_belem QPU.}
\label{fig_belem}
\end{figure}

In order to explore such possibilities, we now test our quantum surrogate model on an existing quantum computer. For this, we choose the five qubit quantum computer ibmq\_belem, as it offers sufficient resources for running our circuits, i.e., availability, number of qubits and gate fidelities. Despite privileged access and using the Qiskit Runtime environment (which does not require re-queuing for each optimization iteration), our multi-layer approach faced difficulties for all higher dimensional functions. However, after a manual hyperparameter search, we were able to obtain a fit for the one-dimensional Griewank function with three qubits, six layers and 100 optimization iterations (executed on the QPU) as depicted in figure~\ref{fig_belem}. While the achieved R2 score of 0.54 is fairly moderate, we observe that in both evaluations (see figure~\ref{fig_6surfaces} and~\ref{fig_belem}), the QNN is very accurate in determining the shape of the underlying function. This can already be valuable information in practice, as it allows for discerning desirable from undesirable subspaces.

\subsection{Analysis of scaling behavior}
Neglecting decoherence, there are three main types of errors on real hardware causing failures: Single-qubit errors, two-qubit gate errors and readout errors. For the ibmq\_belem, the mean Pauli-$X$ error is around 0.04\%, the mean CNOT error roughly amounts to 1.08\% and the readout error is about 2.17\%. Taking the mean Pauli-$X$ error as proxy for single-qubit errors and the mean CNOT error as proxy for two-qubit gate errors, one can approximate the probability of a circuit remaining without error. Subtracting the error rate from one results in the ``survival rate" per gate, which will then be exponentiated with the number of respective gates in the circuit. Our three qubit, six layer circuit for the ibmq\_belem consists (on average\footnote{We are mixing two different circuits in our ansatz, see figure \ref{fig_vqc_example}.}) of 10 single-qubit gates, two two-qubit gates per layer and three readouts, resulting in a total survival rate of 80.13\% for six layers. The survival rates for different circuit sizes with regard to the number of qubits and the number of layers is shown in table~\ref{table: survival_rates}.

Taking this total survival rate of 80.13\% as the minimum requirement for a successful real hardware run, one can also approximate the required error rate improvement such that the four qubits, 20 layer circuit we use for finding a surrogate model for the Griewank function with two-dimensional input data could be run on real hardware. In order to have only one variable to solve for, we keep the single-qubit error constant at its current rate and assume that the ratio of readout error to two-qubit gate error remains at two. This results in a required two-qubit gate error of 0.15\% and a readout error of 0.3\%, resembling a reduction to about 14\% of current error levels. Looking towards more recent IBM QPUs like the Falcon r5.11, lower error rates than the here employed ibmq\_belem are already available: 0.9\% for CNOTs and 0.02\% for Pauli-$X$ gates. For this QPU, our calculations yield a survival rate of 52.76\%, which displays significant improvement, but still prevents modelling the Griewank function sufficiently well.

Note, that this analysis does assume, that the training is executed in an noise-aware manner, i.e., each circuit execution during the training should be repeated multiple times to ensure identifying an error free result by employing a suitable postprocessing routine. As the here demanded success probability of $\sim$80\% is already quite high, this ensures a negligibly small, constant, overhead in computation time.

\begin{table}[t]
\caption{Estimated survival rates on noisy QPUs.}
\centering
\tabcolsep=0.15cm
\renewcommand{\arraystretch}{1.5}
\begin{tabular}{cc|c c c c c|}
\cline{3-7}
& & \multicolumn{5}{ c| }{Number of layers} \\ \cline{3-7}
& & 4 & 8 & 12 & 16 & 20 \\ \cline{1-7}
\multicolumn{1}{ |c  }{\multirow{5}{*}{\shortstack{Number \\ of \\ qubits}} } &
\multicolumn{1}{ |c| }{2} & 0.91 & 0.86 & 0.82 & 0.77 & 0.73\\ 
\multicolumn{1}{ |c  }{}                        &
\multicolumn{1}{ |c| }{4} & 0.79 & 0.67 & 0.58 & 0.49 & 0.42   \\ 
\multicolumn{1}{ |c  }{}                        &
\multicolumn{1}{ |c| }{8} & 0.59 & 0.41 & 0.29 & 0.20 & 0.14    \\ 
\multicolumn{1}{ |c  }{}                        &
\multicolumn{1}{ |c| }{16} & 0.33 & 0.16 & 0.07 & 0.03 & 0.02     \\ 
\multicolumn{1}{ |c  }{}                        &
\multicolumn{1}{ |c| }{32} & 0.10 & 0.02 & 0.00 & 0.00 & 0.00       \\
\cline{1-7}
\end{tabular}
\label{table: survival_rates}
\end{table}



\section{\uppercase{Conclusion}}
\label{sec:conclusion}
Surrogate models have provided enormous cost and time savings in industrial development processes. Nevertheless, dealing with small and noisy data sets still remains a challenge, mostly due to overfitting tendencies of the current state of the art machine learning approaches applied. In this paper, we have demonstrated that quantum surrogate models based on QNNs can offer an advantage over similarly sized classical ANNs in terms of prediction accuracy for substantially more difficult data sets than those used in previous literature, when the sample size is scarce and substantial noise is present. For this, we constructed suitable QNNs, while having employed state-of-the-art ansatz design knowledge, namely: data preprocessing in form of scaling, data reuploading, parallel encoding, layering with a sufficient number of parameters and using different ansätze in one circuit. 



In addition to that, our noise and scaling analyses on quantum surrogate models for higher dimensional inputs, combined with the envisaged reduction of quantum error rates by quantum hardware manufacturers show that our simulation results could be replicated on QPUs in the near future. A possible way to accelerate this process might be switching from a data-reuploading-heavy circuit to one focused on parallel encoding, as this would shorten the overall circuit, allowing for the use of more qubits. Finally, we encourage future work to expand the here indicated trade-off between the solution quality and the number of parameters to also include an analysis of the runtime. This will be especially interesting for increasingly challenging datasets, as current NISQ-restrictions only allow for the exploration of problem instances, that can rapidly be solved by naive classical approaches.



\section*{\uppercase{Acknowledgements}} 
This paper was partially funded by the German Federal Ministry for Economic Affairs and Climate Action through the funding program "Quantum Computing -- Applications for the industry" based on the allowance "Development of digital technologies" (contract number: 01MQ22008A).

\bibliographystyle{apalike}
{\small
\bibliography{main}}


\end{document}